\documentclass[12pt]{article}

\usepackage{geometry}
\usepackage{graphicx}
\usepackage{amssymb}
\usepackage{epstopdf}
\usepackage{subfig}
\usepackage{fullpage}
\usepackage{listings}
\usepackage[super,sort&compress]{natbib}

\usepackage[]{authblk}
\usepackage{booktabs}
\usepackage{multirow}

\usepackage{setspace}

\title{A close halo of large transparent grains around extreme red giant stars}

\date{}                                     

\author[1]{Barnaby R. M. Norris}
\author[1]{Peter G. Tuthill}
\author[1,2,5]{Michael J. Ireland}
\author[3]{Sylvestre Lacour}
\author[4]{Albert A. Zijlstra}
\author[4]{Foteini Lykou}
\author[1,6]{Thomas M. Evans}
\author[1]{Paul Stewart}
\author[1]{Timothy R. Bedding}

\affil[1]{Sydney Institute for Astronomy (SIfA), School of Physics,
University of Sydney, NSW 2006, Australia}
\affil[2]{Department of Physics and Astronomy, Macquarie University,
NSW 2109, Australia}
\affil[3]{LESIA-Observatoire de Paris, CNRS, Universit\'e Pierre et Marie Curie, Universit\'e Paris-Diderot, Meudon, France}
\affil[4]{Jodrell Bank Centre for Astrophysics, School of Physics \&
Astronomy, University of Manchester, Oxford Road, Manchester M13 9PL, UK}
\affil[5]{Australian Astronomical Observatory, PO Box 296, Epping, NSW
1710, Australia}
\affil[6]{Department of Physics, Denys Wilkinson Building, Keble Road, University of Oxford, OX1 3RH, UK}

\begin{document}
\maketitle

\newpage

{\bf
Intermediate-mass stars end their lives by ejecting the bulk of their envelope via a slow dense wind back into the interstellar medium, to form the next generation of stars and planets\citep{Wood1979,Habing1996,Marengo2009}. 
Stellar pulsations are thought to elevate gas to an altitude cool enough for the condensation of dust\citep{Wood1979}, which is then accelerated by radiation pressure from starlight, entraining the gas and driving the wind\citep{Habing1996,Gustafsson2004,Hoefner2009}. 
However accounting for the mass loss has been a problem due to the difficulty in observing tenuous gas and dust tens of milliarcseconds from the star, and there is accordingly no consensus on the way sufficient momentum is transferred from the starlight to the outflow.
Here, we present spatially-resolved, multi-wavelength observations of circumstellar dust shells of three stars on the asymptotic giant branch of the HR diagram.
When imaged in scattered light, dust shells were found at remarkably small radii ($\lesssim$~2 stellar radii) and with unexpectedly large grains ($\sim$300~nm radius).
This proximity to the photosphere argues for dust species that are transparent to starlight and therefore resistant to sublimation by the intense radiation field. 
While transparency usually implies insufficient radiative pressure to drive a wind\citep{Woitke2006,Ireland2006}, the radiation field can accelerate these large grains via photon scattering rather than absorption\citep{Hoefner2008} - a plausible mass-loss mechanism for  lower-amplitude pulsating stars.
}

\bigskip

We observed W Hya, R Dor and R Leo using aperture-masked\citep{Tuthill2000,Lacour2011} polarimetric interferometry (see Figure 1), along with dust-free stars to verify our detection methodology. 
Figure~1 shows the ratio of the horizontally to vertically polarised visibilities ($V_{\rm horiz}/V_{\rm vert}$), plotted as a function of baseline azimuth and length. The dust-free star 2~Cen, which has no polarised flux, shows a constant ratio $V_{\rm horiz}/V_{\rm vert} = 1.0$ within uncertainties. However the dust-enshrouded star W~Hya shows a strong sinusoidal variation of $V_{\rm horiz}/V_{\rm vert}$ with azimuth, as expected from a resolved circumstellar scattering shell. By taking advantage of spherical symmetry, the much simpler baseline-dependent observable plotted in Figure~2 was produced. 
A model was then fitted to the data to determine the dust shell radius and amount of light scattered by the shell at each wavelength (see Supplementary Information). These results are summarised in Table \ref{resultstable}. Figure~3 shows the model image of the star and shell as seen in orthogonal polarisations for W Hya at 1.24$\mu$m, from which the model visibilities were derived.

Scattered-light dust shells around the three AGB stars observed were found close to the star, with radii $\lesssim$2~R$_{\rm star}$. This is in contrast to earlier models which place the shell at many stellar radii\citep{Habing1996}, but is consistent with some recent models\citep{Woitke2006} and with interferometric\citep{Wittkowski2011} and polarimetric\citep{Ireland2005} measurements. 
Based on typical elemental abundances and spectral observations, the composition of AGB dust shells is expected to be dominated by silicates\citep{Waters1996,Gail2003,Hoefner2009}, in the form of olivine (Mg$_{2x}$Fe$_{2(1-x)}$SiO$_4$) and/or pyroxene (Mg$_{x}$Fe$_{(1-x)}$SiO$_3$) where $0 \leq x \leq 1$. The temperature of a grain is determined by its opacity - i.e. how strongly it absorbs the surrounding radiation field. Non-grey models\citep{Woitke2006} show that silicate dust which contains iron absorbs the stellar flux strongly (having high opacities at wavelengths $\sim1\mu m$ where the energy distribution peaks) and so can only condense at distances greater than $\sim 5$ stellar radii. These iron-rich species could be accelerated via absorption of stellar radiation, but they form too far from the star to provide an efficient mass-loss mechanism for low-amplitude pulsators (semi-regular variables)\citep{Woitke2006}.
Our detection of dust much closer to the star is instead consistent with the presence of iron-free silicates such as forsterite (Mg$_2$SiO$_4$) and enstatite (MgSiO$_3$) which are almost transparent at wavelengths of $\sim1\mu$m. Such grains do not heat to sublimation, despite the intense radiation close to the star, but the same transparency also prevents the momentum transfer from starlight required to drive a wind. For some stars, a possible solution to this dilemma arises when very large grains are considered. 

The degree of scattering by dust depends strongly on wavelength and on the size of the particles. By analysing our multi-wavelength measurements using Mie scattering theory, the effective grain size and number of grains were determined. As shown in Figure~4, an effective grain radius of $\sim$300nm was found. For grains of this size, the scattering opacity becomes very large, well beyond that arising from Rayleigh scattering when smaller particles are assumed. In this regime, the contribution to radiative acceleration via scattering, rather than by absorption alone, must be considered. Models show that grains exceeding a certain critical scattering opacity can drive a wind at high Mg condensation and that, for a star of temperature 2700 K, this critical scattering opacity is only exceeded in a narrow range of dust grain radii around $300$ nm\citep{Hoefner2008}. It should also be noted that a narrow range of grain sizes, of the order of $\sim$500nm radius, is predicted based on a self-regulating feedback mechanism: grain growth effectively halts once the critical size is reached, since the dust is then accelerated outwards and gas densities quickly decrease\citep{Hoefner2008}. This is consistent with observations of grains in the inter-stellar medium, which are dominated by silicates\citep{Draine2003} and have similar grain sizes\citep{Frisch1999,Weingartner2001}.
Wind driving due to scattering by Mg-rich silicates is consistent with the finding that mass-loss in these stars depends on their metallicity\citep{Lagadec2008}. 
While this model encounters difficulties in the case of stars with extremely extended atmospheres, such as R Leo (due to the mass of the stellar atmosphere at and above the dust-forming layers being too high to allow sufficient acceleration\citep{Ireland2011}), it provides a plausible explanation for the mass loss of semi-regular pulsating stars such as R~Dor. 
Our observations provide direct evidence for a population of dust grains capable of powering a scattering-driven wind.

The last column of Table \ref{resultstable} gives the mass of the dust that contributes to the observed scattering signal, assuming the shell to be thin and the dust grains to be forsterite of a uniform size. If full Mg condensation and solar abundances are assumed then the gas-to-dust ratio is $\sim$600, yielding total shell masses of $\sim6\times10^{-7}$ M$_\odot$, $\sim2\times10^{-7}$ M$_\odot$ and $\sim1\times10^{-7}$ M$_\odot$ for W Hya, R Dor and R Leo, respectively.
Since the pulsation periods of these stars are $\sim$1 year and the mass loss rates are $\sim1\times10^{-7}$ M$_\odot$/year\citep{Loup1993,Olofsson2002}, this implies that for stars with less extended atmospheres a significant fraction of the observed shell is ejected each pulsation cycle, consistent with the observed dust being part of an outflow. In the extended-atmosphere case, where full Mg condensation is not observationally supported, a possible alternative dust species is corundum (Al$_2$O$_3$), as discussed in the Supplementary Information. Further time-dependent grain growth and dynamical models will help elucidate the role played by light scattered from large-grained dust in the AGB star mass-loss process.

\newpage

\bigskip

\begin{table}[htbp]\scriptsize
  \centering
  \begin{tabular}{@{} cccccccc @{}}
    \toprule
  Star & $\Phi$ & $\lambda$ ($\mu$m) & $R_{\rm star}$ (mas) & $R_{\rm shell}$ (mas) & Scattered fraction & Grain radius & Scattering-shell mass  \\
    \midrule
    
     {R Dor}  & 0.7   & 1.04 & 27.2$\pm$0.2 & 43.3$\pm$0.3 & 0.124$\pm$0.003 & {299$\pm$39 nm} & {$2.7\pm0.2\times10^{-10}$ M$_\odot$}  \\
                 &    & 2.06 & 27.7$\pm$1.4 & 43.6$\pm$3.2 & 0.014$\pm$0.002 & ¥ & ¥ \\ [1ex]
     {W Hya} &  0.2  & 1.04 & 18.7$\pm$0.4 & 37.9$\pm$0.2 & 0.176$\pm$0.002 & {316$\pm$4 nm} & {$1.04\pm0.02\times10^{-9}$ M$_\odot$}  \\
                 &    & 1.24 & 18.9$\pm$0.5 & 37.0$\pm$0.3 & 0.110$\pm$0.003 & ¥ & ¥  \\
    ¥            &    & 2.06 & 18.9 (fixed) & 37.0 (fixed) & 0.022$\pm$0.004 & ¥ & ¥  \\ [1ex]
    R Leo    &  0.4   & 1.04 & 18.3$\pm$0.3 & 29.9$\pm$0.4 & 0.120$\pm$0.003 &$\sim$300 nm$^\dag$ & $\sim2\times10^{-10}$ M$_\odot$ \\
       
    \bottomrule
  \end{tabular}
  \caption{{\bf Summary of fitted model parameters.} The radii of the dust shells were found to be $\lesssim2R_{\rm star}$. The scattered fraction is the proportion of the total flux arising from scattering by the dust shell. The grain radius was obtained from fitting to multi-wavelength observations using Mie scattering (value for R~Leo fixed at 300nm). The scattering-shell mass was calculated only for the observed dust (and not, for example, a distribution extending to small grains invisible to our technique). Stellar radii given are for a uniform disc. All three AGB stars - W Hya, R Dor and R Leo - were observed in March 2009, with additional observations of W Hya at 1.24$\mu$m and 2.06$\mu$m made in June 2010. Although the photospheric diameter - and possibly the dust shell diameter - are expected to vary throughout the stellar pulsation cycle, the two sets of observations for W Hya were taken approximately one period apart, and so have been combined. These figures assume the dust to be an iron-free silicate (forsterite) grains of uniform size, and full Mg condensation. Full Mg condensation is a reasonable assumption for the stars with more compact atmospheres (e.g. R Dor) but is inconsistent with observed optical depths for stars with more extended atmospheres\citep{Ireland2011}. In the event that there is also a population of small weakly scattering grains which do not show up in our data, these values represent lower limits with the total shell mass being higher. Furthermore, if the shell is geometrically extended then the true mass will be higher, as these calculations assume a thin shell. The uncertainties given are based on random errors and do not account for systematic errors such as those described here.
Hipparcos parallaxes\citep{VanLeeuwen2007} and experimentally measured optical constants\citep{Jager2003} have been used. $\Phi$ indicates visual phase, derived from the AAVSO International Database. $^\dag$Assumed grain size.}
  \label{resultstable}
\end{table}

\newpage

\begin{figure}[h]
  \centering
  \includegraphics[]{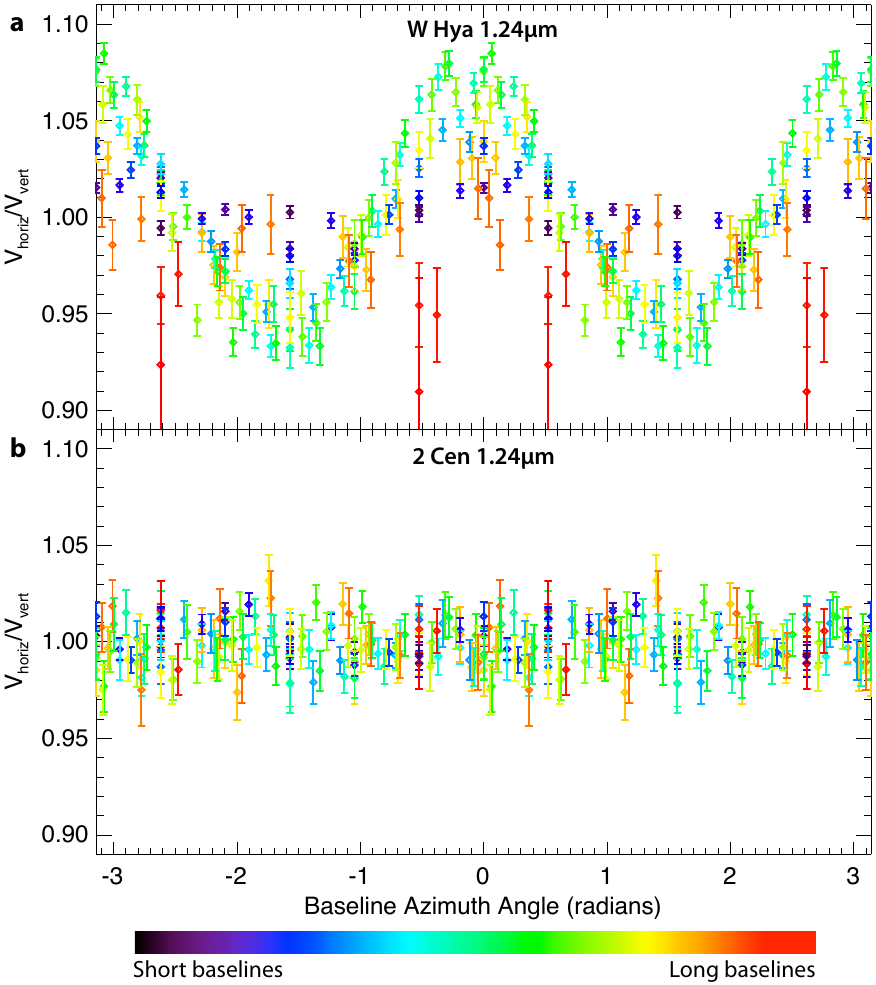}
  \caption{{\bf Polarimetric interferometry of W Hya at 1.24$\mu$m} - 
Although light scattered by each part of the circumstellar dust shell is strongly polarised, the integrated signal recovered with conventional polarimetry is zero for an unresolved spherically symmetric shell. In this study, aperture masking interferometry\citep{Tuthill2000,Lacour2011} (which converts the 8~m VLT pupil into a multi-element interferometer, using the NACO\citep{Lenzen2003} instrument) allows access to the $\sim$10 mas spatial scales required to resolve the shell, while polarimetric measurements (obtained by simultaneously measuring interferometric visibilities in orthogonal polarisations\citep{Tuthill2010}) allows light from the star and from the dust shell to be differentiated.
Here, the ratio of the horizontally polarised visibilities ($V_{\rm horiz}$) to vertically polarised visibilities ($V_{\rm vert}$) is plotted against baseline azimuth angle (corresponding to position angle on the sky). Colour encodes the baseline length (the longest being 7.3~m). $V_{\rm horiz}/V_{\rm vert}$ is a \emph{differential} observable, which allows the cancellation of residual systematic errors and depends only on the fractional polarised scattered light signal.
Panel a) shows the result for W~Hya, an AGB star with a circumstellar shell; $V_{\rm horiz}/V_{\rm vert}$ deviates from 1, varying sinusoidally with azimuth. This is the signal expected from a thin, spherically symmetric dust shell scattering the light from a central star. Different baselines exhibit varying amplitudes of this signal, encoding the spatial extent of the resolved structure. Data points have been repeated over 2 cycles. Error bars are 1$\sigma$. The longest baselines (red) have poor signal-to-noise because they are close to the null where the visibility curve of the star is extremely low. Panel b) shows visibility data for the star 2~Cen, which has no circumstellar dust shell and hence no polarised signal from scattering; here, the ratio $V_{\rm horiz}/V_{\rm vert}$ is $\sim$1 for all azimuths.}
  \label{vhvvplots}
\end{figure}

\newpage

\begin{figure}[h]
  \centering
  \includegraphics[]{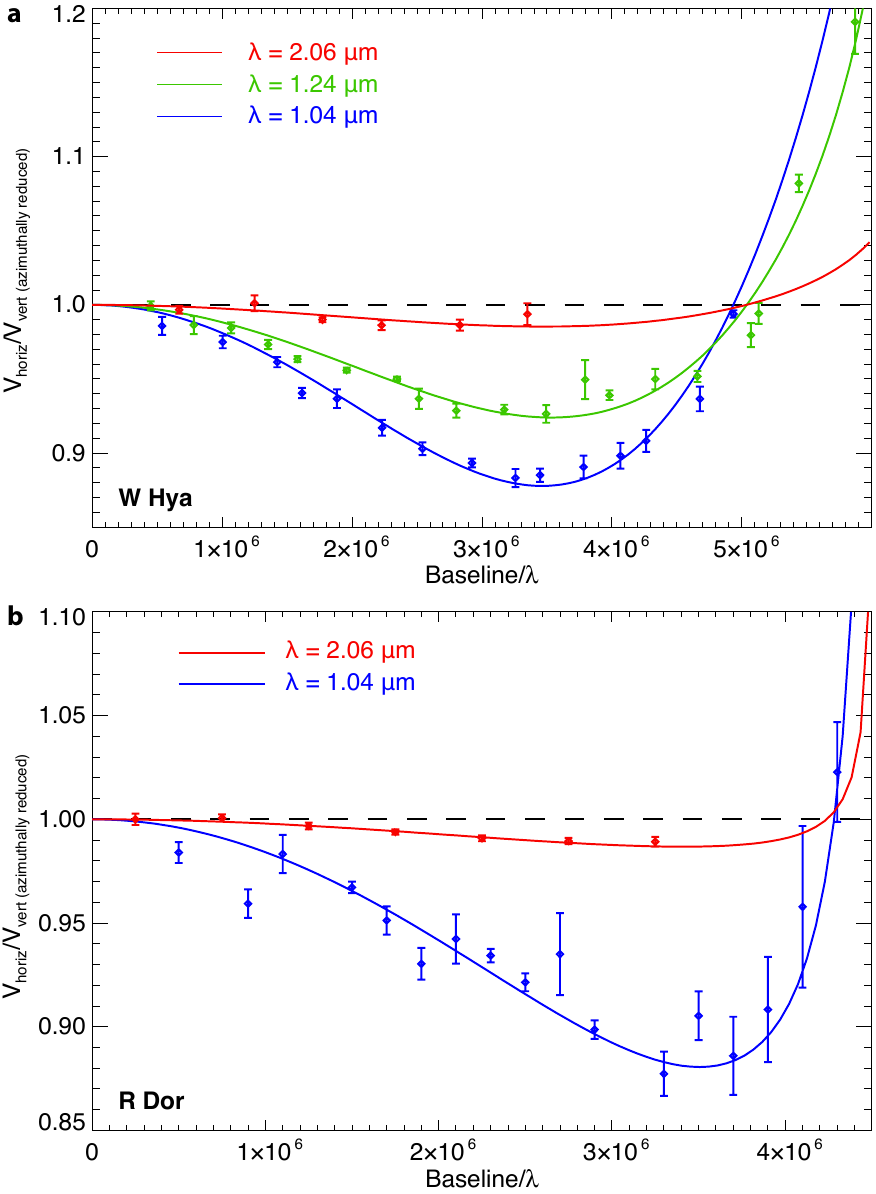}
  \caption{{\bf Wavelength dependence of scattering for W~Hya and R~Dor} - The azimuthally reduced $V_{\rm horiz}/V_{\rm vert}$ for the AGB stars W Hya and R Dor, plotted against spatial frequency (baseline length/$\lambda$), at multiple wavelengths. The functional form of the visibility ratio excursions exhibited by all three AGB stars are consistent, within uncertainties, with a simple spherically-symmetric shell. We were therefore able to significantly enhance our signal-to-noise in a quantitative analysis by reducing our two-dimensional visibility data to a one-dimensional function of the baseline length (corresponding to spatial frequency). This was achieved by dividing $V_{\rm horiz}/V_{\rm vert}$ by the expected sinusoidal variation (characteristic of a spherical shell) at a fixed amplitude, resulting in the much simpler baseline-dependent observable plotted here. Points with 1$\sigma$ error bars are the observed data (binned) and the solid lines are the fitted model. The spatial frequency of the characteristic `dip' varies with the radius of the dust shell: for a larger shell the dip occurs at lower spatial frequencies. The depth of the dip depends on the amount of scattered light, with a larger deviation from $V_{\rm horiz}/V_{\rm vert} = 1.0$ indicating that a larger fraction of the total flux arises from light scattered by the shell. This is seen to decrease strongly at longer wavelengths as expected theoretically; the precise change in this quantity as a function of wavelength can be used to determine the dust grain radius using Mie scattering theory (see Figure~4). The fitted parameters for these quantities are included in Table \ref{resultstable}. The 2.06$\mu$m data have insufficient spatial resolution to constrain the shell size, so for models at these wavelengths the shell size was fixed to be consistent with the fitted size at shorter wavelengths.}
  \label{vhvvplots}
\end{figure}

\begin{figure}[h]
  \centering
  \includegraphics[]{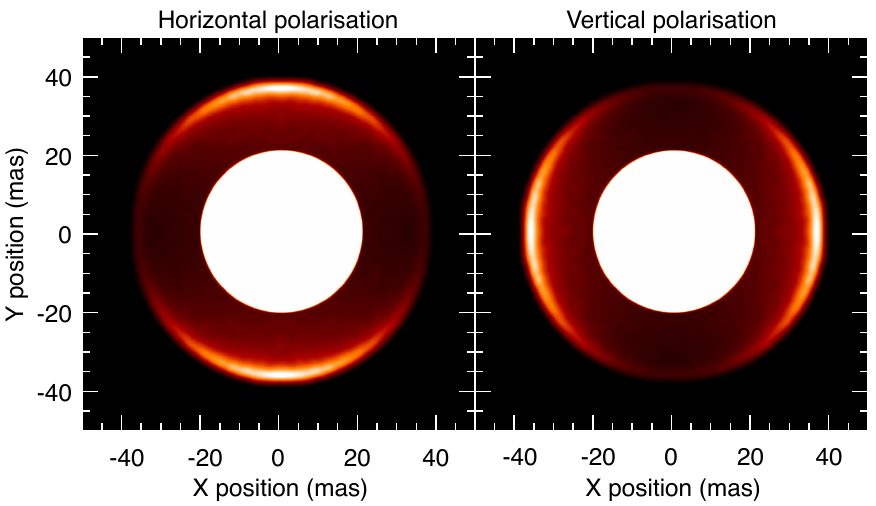}
  \caption{{\bf Model image for W Hya with circumstellar shell viewed in horizontally and vertically polarised light} - The white disc represents the uniform-disc star used in the model. A three-dimensional model of a star with a thin scattering shell was constructed, and the scattered intensity observed in each polarisation for each point on the shell was calculated using Mie scattering, finally yielding an image of the star and shell. Polarised visibilities were then derived from the model and fitted to the observed visibilities, to allow dust shell radius and scattered fraction to be determined. Details of the modelling process can be found in the supplementary information. Refer to Figure~S2 for a diagram illustrating how the polarised intensity distribution arises.}
  \label{vhvvplots}
\end{figure}

\begin{figure}[h]
  \centering
  \includegraphics[]{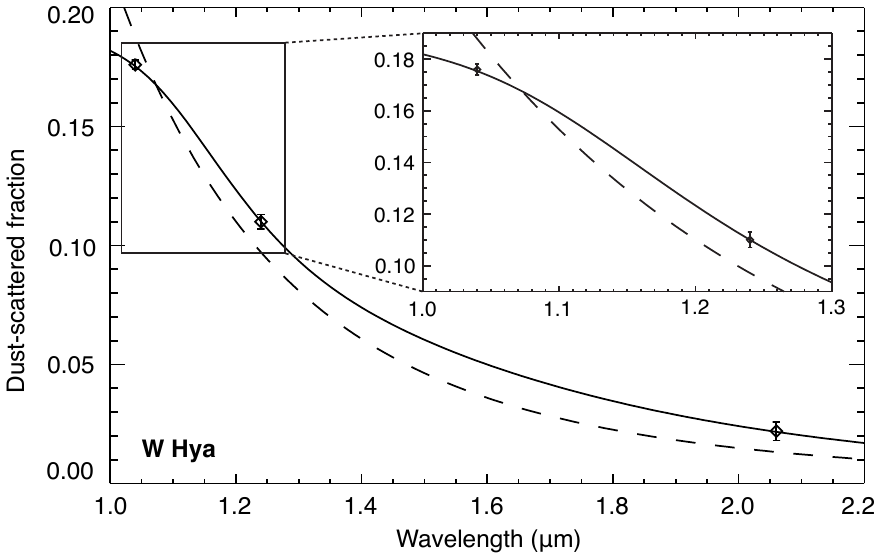}
  \caption{{\bf Grain size measurement} - Grain-size fitted to the scattered light fraction as a function of wavelength, for W Hya. Error bars are 1$\sigma$. The inset gives a blowup of the 1.0--1.3\,$\mu$m region. The solid line represents the fitted Mie scattering model (where grain size and grain number were fitted parameters), while the dashed line represents the best Rayleigh fit (grain size fixed to be below the Rayleigh limit). The data are inconsistent with Rayleigh scattering. The fitted Mie model yields an effective grain radius of 316$\pm$4 nm. In reality a distribution of grain sizes may be present, for example a population of very much smaller particles would contribute only weakly and could lie undetected. Our data show that, regardless of the presence or absence of smaller grains, a population of large $\sim$300nm grains is required. }
  \label{vhvvplots}
\end{figure}

\clearpage
\newpage

\section*{Supplementary Information}
\subsection*{Modelling Details}
To determine the size of the dust shell and the proportion of the total flux arising from scattering by the dust shell, a model was created and fitted to the observed differential visibilities. The star was modelled as a uniform disk. While AGB stars may generally be better represented as a limb-darkened disk, it was found that using a limb-darkened disk model made a negligible difference to the fitted parameters describing the dust shell. Thus for the sake of simplicity, and to allow unambiguous reference to previously measured uniform-disk diameters, a uniform disk was used here. The assumption was imposed that the dust shell is of negligible thickness compared to its radius, and so was modelled as a thin spherically symmetric shell. While in reality the shell may be thick or form the inner edge of an outflow, due to the rapid fall-off with radius of both illumination and density (in the case of an outflow) the polarimetric signal is dominated by the bright inner surface, and hence our data is unable to constrain a more complex density profile. Furthermore, we are also unable to resolve the thickness of the shell. If the shell is significantly extended, then the true mass of the shell will be higher than the ``scattering shell mass'' estimated here.

Next, images of the shell as seen in linearly polarised light were created. The following description refers to the angles labelled in Figure~S1. The polarised intensity of light scattered by a dust grain towards an observer is a function of the angle between the ray incident upon the grain and the polarisation vector measured by the observer $\alpha$, as well as wavelength, dust-grain size and refractive index of the material. Due to the close proximity of the dust to the star, the photosphere has a large angular size as seen from the point of view of a dust grain (represented by $\theta$ and $\phi$). Hence for each dust grain the polarised intensity of the scattered light seen by the observer was integrated over $\theta$ and $\phi$, i.e. over the entire visible disc of the star. This scattered intensity was calculated using the University of Oxford Mie scattering code\citep{Oxford2010}. It is assumed in this model that the shell is optically thin and multiple scattering does not occur. This calculation was repeated for each point in the dust shell in three dimensions, and then integrated along the line-of-sight of the observer to form a set of two-dimensional images of the shell. Once the image of the star was added to the image of the dust, a Fourier transform of each polarised image was used to produce the polarised visibilities of the star and shell, which could then be fitted to the observed data.

\subsection*{Discussion}
In the extended-atmosphere case, where full Mg condensation is not supported by observations, a possible alternative dust species to iron-free silicates is corundum (Al$_2$O$_3$), which is also largely transparent to the stellar flux. While the overall dust composition of AGB stars may be dominated by silicates, corundum may play an important role in dust formation and in radiative transfer processes\citep{Maldoni2005}. Gas temperatures at the observed dust shell radii derived from stellar models are low enough for both iron-free silicates and corundum to form (depending on the stellar phase)\citep{Schmid1981,Ireland2006}. Corundum generally forms earlier in the dust condensation sequence than silicates\citep{Lorenz2000} and may form a condensation site around which a silicate mantle later condenses\citep{Kozasa1997}. If the dust observed here is corundum then this is consistent with other evidence for corundum existing as close as 1.5 stellar radii around AGB stars\citep{Verhoelst2006,Wittkowski2011}. The low abundance of Al means full condensation would yield an optical depth consistent with our data, and if the observed scattering shell is assumed to be entirely composed of corundum then a total shell mass $\sim$20 times that given above is implied. In reality, the role of different dust species in the shell may be more complex.

\bigskip
\bigskip

\begin{figure}[h]
  \centering
  \includegraphics[width=9cm]{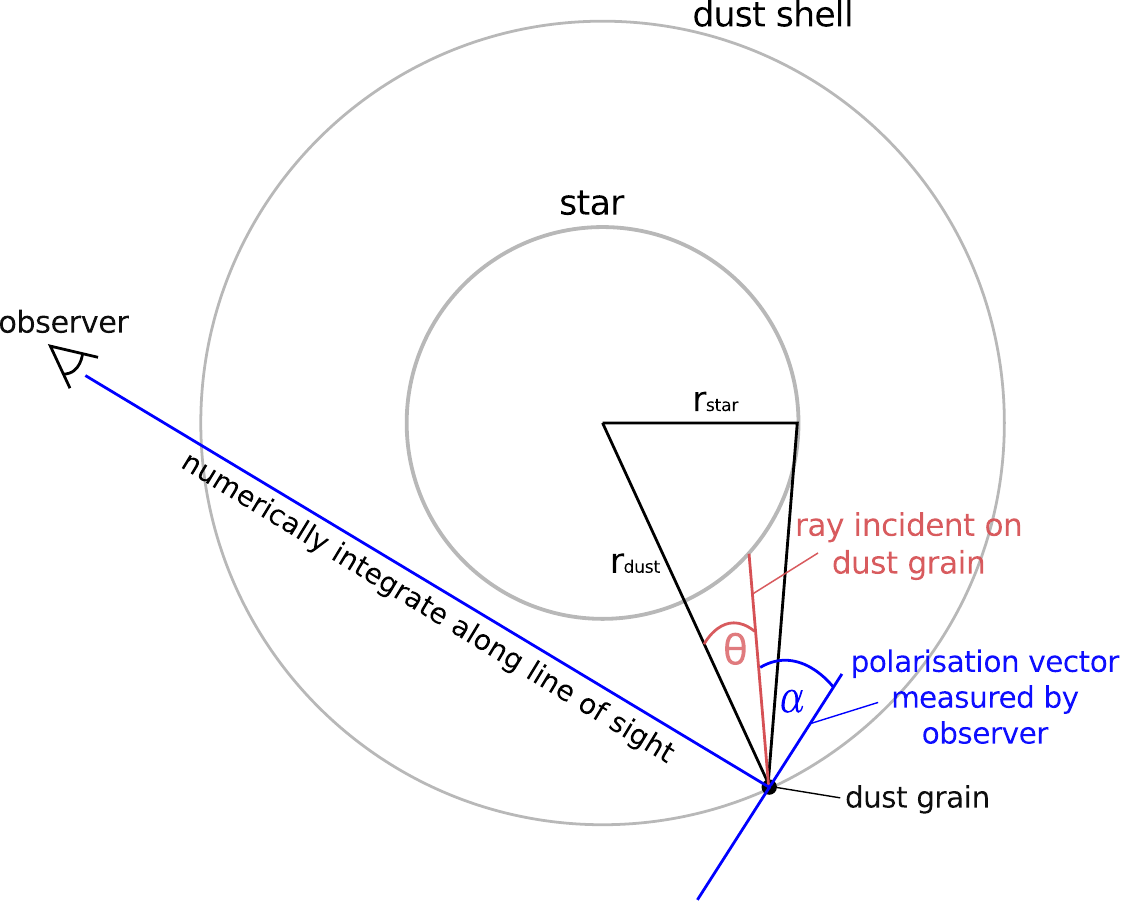}
  \caption*{{\bf Figure S1: }A diagram of the numerical model used to calculate the polarised intensities of scattered light. A description of the modelling procedure is included in the text of the supplementary information. The angle $\phi$ is not shown for clarity. }
  \label{modeldiagram}
\end{figure}

\bigskip
\bigskip
\bigskip
\bigskip

\begin{figure}[h]
  \centering
  \includegraphics[]{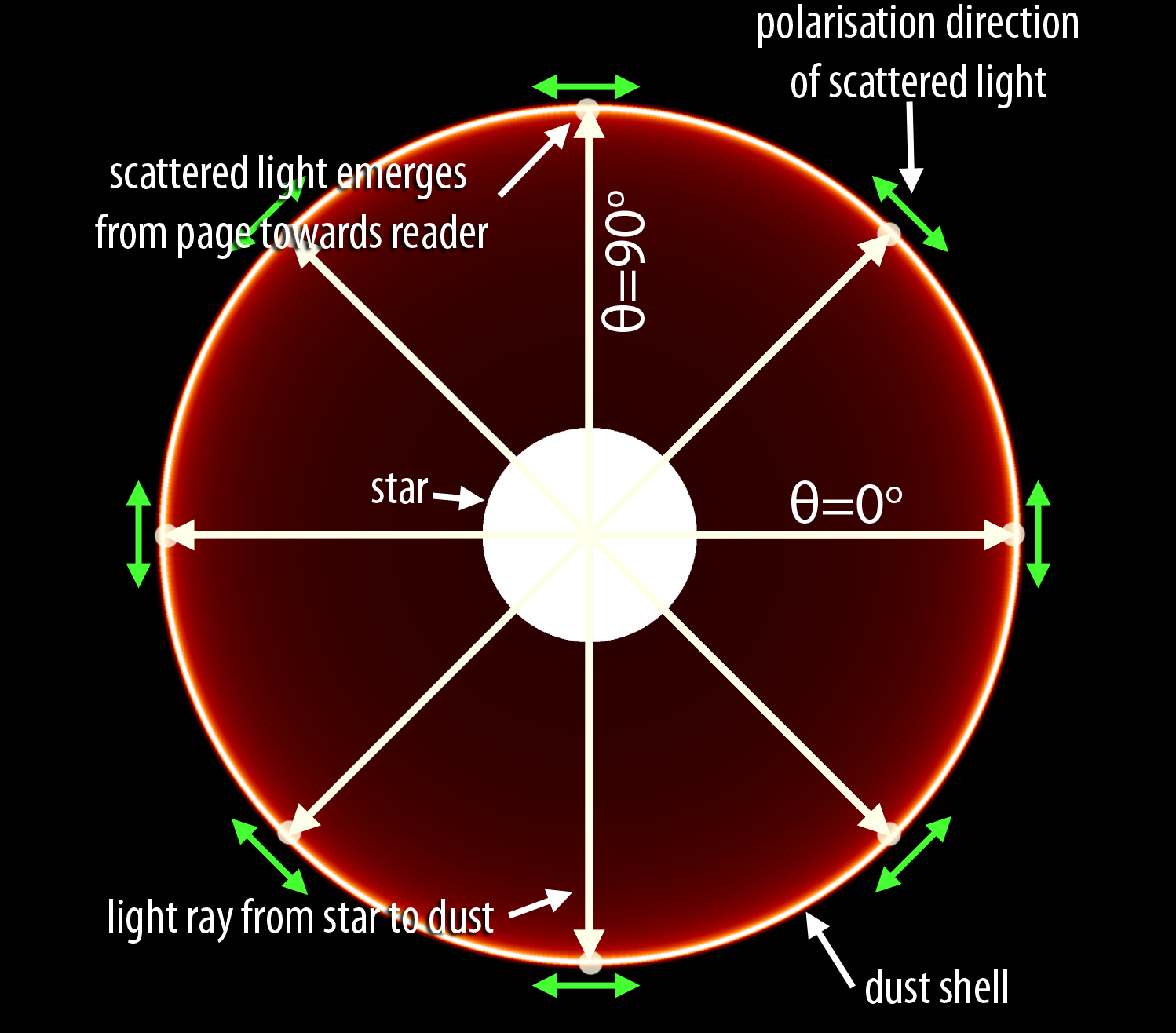}
  \caption*{{\bf Figure S2: }A diagram showing how the polarisation of starlight by a spherical dust shell arises. Unpolarised light emerges from the star and is scattered by dust in the spherical shell in the direction out of the page. This scattered light is linearly polarised, such that its polarisation is in a direction perpendicular to the plane defined by the light ray incident on the dust grain and the ray going from the dust grain to the observer. In this way, if the shell were viewed through a horizontal polariser, light reaching the viewer from the shell at $\theta = 90^\circ$  will have its full intensity while light coming from the shell at $\theta = 0^\circ$ will have zero intensity (compare to Figure 3).}
  \label{shelldiag}
\end{figure}

\clearpage
\newpage


\begin{thebibliography}{30}
\providecommand{\natexlab}[1]{#1}
\providecommand{\url}[1]{\texttt{#1}}
\expandafter\ifx\csname urlstyle\endcsname\relax
  \providecommand{\doi}[1]{doi: #1}\else
  \providecommand{\doi}{doi: \begingroup \urlstyle{rm}\Url}\fi

\bibitem[{Wood}(1979)]{Wood1979}
P.~R. {Wood}.
\newblock {Pulsation and mass loss in Mira variables}.
\newblock \emph{Astrophys. J.}, 227:\penalty0 220--231 (1979)

\bibitem[{Habing}(1996)]{Habing1996}
H.~J. {Habing}.
\newblock {Circumstellar envelopes and Asymptotic Giant Branch stars}.
\newblock \emph{Astron. Astrophys. Rev.}, 7:\penalty0 97--207 (1996)

\bibitem[{Marengo}(2009)]{Marengo2009}
M.~{Marengo}.
\newblock {A Review of AGB Mass Loss Imaging Techniques}.
\newblock \emph{Proceedings of the Astronomical Society of Australia},
  26:\penalty0 365--371 (2009)

\bibitem[Gustafsson and Hoefner(2004)]{Gustafsson2004}
B.~Gustafsson and S.~Hoefner.
\newblock \emph{Asymptotic Giant Branch Stars, H. J. Habing and H. Olofsson,
  eds.}, chapter 4 - Atmospheres of AGB Stars.
\newblock Springer (2004)

\bibitem[{H{\"o}fner}(2009)]{Hoefner2009}
S.~{H{\"o}fner}.
\newblock {Dust Formation and Winds around Evolved Stars: The Good, the Bad and
  the Ugly Cases}.
\newblock In {T.~Henning, E.~Gr{\"u}n, \& J.~Steinacker}, editor,
 volume 414 of
  \emph{Astronomical Society of the Pacific Conference Series} (2009)

\bibitem[{Woitke}(2006)]{Woitke2006}
P.~{Woitke}.
\newblock {Too little radiation pressure on dust in the winds of oxygen-rich
  AGB stars}.
\newblock \emph{Astron. Astrophys.}, 460:\penalty0 L9--L12 (2006)

\bibitem[{Ireland} and {Scholz}(2006)]{Ireland2006}
M.~J. {Ireland} and M.~{Scholz}.
\newblock {Observable effects of dust formation in dynamic atmospheres of
  M-type Mira variables}.
\newblock \emph{Mon. Not. R. Astron. Soc.}, 367:\penalty0 1585--1593 (2006)

\bibitem[{H{\"o}fner}(2008)]{Hoefner2008}
S.~{H{\"o}fner}.
\newblock {Winds of M-type AGB stars driven by micron-sized grains}.
\newblock \emph{Astron. Astrophys.}, 491:\penalty0 L1--L4 (2008)

\bibitem[{Tuthill} et~al.(2000){Tuthill}, {Monnier}, {Danchi}, {Wishnow}, and
  {Haniff}]{Tuthill2000}
P.~G. {Tuthill}, J.~D. {Monnier}, W.~C. {Danchi}, E.~H. {Wishnow}, and C.~A.
  {Haniff}.
\newblock {Michelson Interferometry with the Keck I Telescope}.
\newblock \emph{Publications of the Astronomical Society of the Pacific}, 112:\penalty0 555--565 (2000)

\bibitem[{Lacour} et~al.(2011){Lacour}, {Tuthill}, {Amico}, {Ireland},
  {Ehrenreich}, {Huelamo}, and {Lagrange}]{Lacour2011}
S.~{Lacour}, P.~{Tuthill}, P.~{Amico}, M.~{Ireland}, D.~{Ehrenreich},
  N.~{Huelamo}, and A.M. {Lagrange}.
\newblock {Sparse aperture masking at the VLT. I. Faint companion detection
  limits for the two debris disk stars HD 92945 and HD 141569}.
\newblock \emph{Astron. Astrophys.}, 532:\penalty0 A72+ (2011)

\bibitem[{Wittkowski} et~al.(2011){Wittkowski}, {Boboltz}, {de Breuck}, {Gray},
  {Humphreys}, {Ireland}, {Karovicova}, {Ohnaka}, {Ruiz-Velasco}, {Scholz},
  {Whitelock}, and {Zijlstra}]{Wittkowski2011}
M.~{Wittkowski}, D.~A. {Boboltz}, C.~{de Breuck}, M.~{Gray}, E.~{Humphreys},
  M.~{Ireland}, I.~{Karovicova}, K.~{Ohnaka}, A.~E. {Ruiz-Velasco},
  M.~{Scholz}, P.~{Whitelock}, and A.~{Zijlstra}.
\newblock {The extended atmospheres of Mira variables probed by VLTI, VLBA, and
  APEX}.
\newblock \emph{Astronomical Society of the Pacific Conference Series},
  445\penalty0 (107) (2011)

\bibitem[{Ireland} et~al.(2005){Ireland}, {Tuthill}, {Davis}, and
  {Tango}]{Ireland2005}
M.~J. {Ireland}, P.~G. {Tuthill}, J.~{Davis}, and W.~{Tango}.
\newblock {Dust scattering in the Miras R Car and RR Sco resolved by optical
  interferometric polarimetry}.
\newblock \emph{Mon. Not. R. Astron. Soc.}, 361:\penalty0 337--344 (2005)

\bibitem[{Waters} et~al.(1996){Waters}, {Molster}, {de Jong}, {Beintema},
  {Waelkens}, {Boogert}, {Boxhoorn}, {de Graauw}, {Drapatz}, {Feuchtgruber},
  {Genzel}, {Helmich}, {Heras}, {Huygen}, {Izumiura}, {Justtanont}, {Kester},
  {Kunze}, {Lahuis}, {Lamers}, {Leech}, {Loup}, {Lutz}, {Morris}, {Price},
  {Roelfsema}, {Salama}, {Schaeidt}, {Tielens}, {Trams}, {Valentijn},
  {Vandenbussche}, {van den Ancker}, {van Dishoeck}, {Van Winckel},
  {Wesselius}, and {Young}]{Waters1996}
L.~B.~F.~M. {Waters}, F.~J. {Molster}, T.~{de Jong}, D.~A. {Beintema},
  C.~{Waelkens}, A.~C.~A. {Boogert}, D.~R. {Boxhoorn}, T.~{de Graauw},
  S.~{Drapatz}, H.~{Feuchtgruber}, R.~{Genzel}, F.~P. {Helmich}, A.~M. {Heras},
  R.~{Huygen}, H.~{Izumiura}, K.~{Justtanont}, D.~J.~M. {Kester}, D.~{Kunze},
  F.~{Lahuis}, H.~J.~G.~L.~M. {Lamers}, K.~J. {Leech}, C.~{Loup}, D.~{Lutz},
  P.~W. {Morris}, S.~D. {Price}, P.~R. {Roelfsema}, A.~{Salama}, S.~G.
  {Schaeidt}, A.~G.~G.~M. {Tielens}, N.~R. {Trams}, E.~A. {Valentijn},
  B.~{Vandenbussche}, M.~E. {van den Ancker}, E.~F. {van Dishoeck}, H.~{Van
  Winckel}, P.~R. {Wesselius}, and E.~T. {Young}.
\newblock {Mineralogy of oxygen-rich dust shells.}
\newblock \emph{Astron. Astrophys.}, 315:\penalty0 L361--L364 (1996)

\bibitem[{Gail}(2003)]{Gail2003}
{H.-P.} {Gail}.
\newblock {Formation and Evolution of Minerals in Accretion Disks and Stellar
  Outflows}.
\newblock In {T.~K.~Henning}, editor, \emph{Astromineralogy}, volume 609 of
  \emph{Lecture Notes in Physics, Berlin Springer Verlag}, pages 55--120 (2003)

\bibitem[{Draine}(2003)]{Draine2003}
B.~T. {Draine}.
\newblock {Interstellar Dust Grains}.
\newblock \emph{Annual Review of Astronomy \& Astrophysics}, 41:\penalty0 241--289 (2003)

\bibitem[{Frisch} et~al.(1999){Frisch}, {Dorschner}, {Geiss}, {Greenberg},
  {Gr{\"u}n}, {Landgraf}, {Hoppe}, {Jones}, {Kr{\"a}tschmer}, {Linde},
  {Morfill}, {Reach}, {Slavin}, {Svestka}, {Witt}, and {Zank}]{Frisch1999}
P.~C. {Frisch}, J.~M. {Dorschner}, J.~{Geiss}, J.~M. {Greenberg},
  E.~{Gr{\"u}n}, M.~{Landgraf}, P.~{Hoppe}, A.~P. {Jones}, W.~{Kr{\"a}tschmer},
  T.~J. {Linde}, G.~E. {Morfill}, W.~{Reach}, J.~D. {Slavin}, J.~{Svestka},
  A.~N. {Witt}, and G.~P. {Zank}.
\newblock {Dust in the Local Interstellar Wind}.
\newblock \emph{Astrophys. J.}, 525:\penalty0 492--516 (1999)

\bibitem[{Weingartner} and {Draine}(2001)]{Weingartner2001}
J.~C. {Weingartner} and B.~T. {Draine}.
\newblock {Dust Grain-Size Distributions and Extinction in the Milky Way, Large
  Magellanic Cloud, and Small Magellanic Cloud}.
\newblock \emph{Astrophys. J.}, 548:\penalty0 296--309 (2001)

\bibitem[{Lagadec} and {Zijlstra}(2008)]{Lagadec2008}
E.~{Lagadec} and A.~A. {Zijlstra}.
\newblock {The trigger of the asymptotic giant branch superwind: the importance
  of carbon}.
\newblock \emph{Mon. Not. R. Astron. Soc.}, 390:\penalty0 L59--L63 (2008)

\bibitem[{Ireland} et~al.(2011){Ireland}, {Scholz}, and {Wood}]{Ireland2011}
M.~J. {Ireland}, M.~{Scholz}, and P.~R. {Wood}.
\newblock {Dynamical opacity-sampling models of Mira variables - II.
  Time-dependent atmospheric structure and observable properties of four M-type
  model series}.
\newblock \emph{Mon. Not. R. Astron. Soc.}, 418:\penalty0 114--128 (2011)

\bibitem[{Loup} et~al.(1993){Loup}, {Forveille}, {Omont}, and {Paul}]{Loup1993}
C.~{Loup}, T.~{Forveille}, A.~{Omont}, and J.~F. {Paul}.
\newblock {CO and HCN observations of circumstellar envelopes. A catalogue -
  Mass loss rates and distributions}.
\newblock \emph{Astronomy and Astrophysics Supplement Series}, 99:\penalty0 291--377 (1993)

\bibitem[{Olofsson} et~al.(2002){Olofsson}, {Gonz{\'a}lez Delgado},
  {Kerschbaum}, and {Sch{\"o}ier}]{Olofsson2002}
H.~{Olofsson}, D.~{Gonz{\'a}lez Delgado}, F.~{Kerschbaum}, and F.~L.
  {Sch{\"o}ier}.
\newblock {Mass loss rates of a sample of irregular and semiregular M-type
  AGB-variables}.
\newblock \emph{Astron. Astrophys.}, 391:\penalty0 1053--1067 (2002)

\bibitem[{van Leeuwen}(2007)]{VanLeeuwen2007}
F.~{van Leeuwen}.
\newblock {Validation of the new Hipparcos reduction}.
\newblock \emph{Astron. Astrophys.}, 474:\penalty0 653--664 (2007)

\bibitem[{J{\"a}ger} et~al.(2003){J{\"a}ger}, {Dorschner}, {Mutschke}, {Posch},
  and {Henning}]{Jager2003}
C.~{J{\"a}ger}, J.~{Dorschner}, H.~{Mutschke}, T.~{Posch}, and T.~{Henning}.
\newblock {Steps toward interstellar silicate mineralogy. VII. Spectral
  properties and crystallization behaviour of magnesium silicates produced by
  the sol-gel method}.
\newblock \emph{Astron. Astrophys.}, 408:\penalty0 193--204 (2003)

\bibitem[{Lenzen} et~al.(2003){Lenzen}, {Hartung}, {Brandner}, {Finger},
  {Hubin}, {Lacombe}, {Lagrange}, {Lehnert}, {Moorwood}, and
  {Mouillet}]{Lenzen2003}
R.~{Lenzen}, M.~{Hartung}, W.~{Brandner}, G.~{Finger}, N.~N. {Hubin},
  F.~{Lacombe}, A.-M. {Lagrange}, M.~D. {Lehnert}, A.~F.~M. {Moorwood}, and
  D.~{Mouillet}.
\newblock {NAOS-CONICA first on sky results in a variety of observing modes}.
\newblock In {M.~Iye \& A.~F.~M.~Moorwood}, editor, \emph{Society of
  Photo-Optical Instrumentation Engineers (SPIE) Conference Series}, volume
  4841 of \emph{Society of Photo-Optical Instrumentation Engineers (SPIE)
  Conference Series}, pages 944--952 (2003)

\bibitem[{Tuthill} et~al.(2010){Tuthill}, {Lacour}, {Amico}, {Ireland},
  {Norris}, {Stewart}, {Evans}, {Kraus}, {Lidman}, {Pompei}, and
  {Kornweibel}]{Tuthill2010}
P.~{Tuthill}, S.~{Lacour}, P.~{Amico}, M.~{Ireland}, B.~{Norris}, P.~{Stewart},
  T.~{Evans}, A.~{Kraus}, C.~{Lidman}, E.~{Pompei}, and N.~{Kornweibel}.
\newblock {Sparse aperture masking (SAM) at NAOS/CONICA on the VLT}.
\newblock In \emph{Society of Photo-Optical Instrumentation Engineers (SPIE)
  Conference Series}, volume 7735 of \emph{Society of Photo-Optical
  Instrumentation Engineers (SPIE) Conference Series} (2010)

\bibitem[{University of Oxford}(2010)]{Oxford2010}
Department of Atmospheric, Oceanic and Planetary Physics, University of Oxford.
\newblock {Light Scattering Routines}.
\newblock \emph{http://www-atm.physics.ox.ac.uk/code/mie/index.html} \penalty0 (2010).

\bibitem[{Maldoni} et~al.(2005){Maldoni}, {Ireland}, {Smith}, and
  {Robinson}]{Maldoni2005}
{Maldoni}, M.~M., {Ireland}, T.~R., {Smith}, R.~G. and {Robinson}, G.
\newblock {Al$_{2}$O$_{3}$ dust in OH/IR stars}.
\newblock \emph{Mon. Not. R. Astron. Soc.} {362}, 872--878 (2005).

\bibitem[{Schmid-Burgk} and {Scholz}(1981)]{Schmid1981}
{Schmid-Burgk}, J. and {Scholz}, M.
\newblock {On the possibility of formation of dust grains in M-type
  photospheres}.
\newblock \emph{Mon. Not. R. Astron. Soc.} {194}, 805--808 (1981).

\bibitem[{Lorenz-Martins} and {Pompeia}(2000)]{Lorenz2000}
{Lorenz-Martins}, S. and {Pompeia}, L.
\newblock {Modelling of oxygen-rich envelopes using corundum and silicate
  grains}.
\newblock \emph{Mon. Not. R. Astron. Soc.} {315}, 856--864 (2000).

\bibitem[{Kozasa} and {Sogawa}(1997)]{Kozasa1997}
{Kozasa}, T. and {Sogawa}, H.
\newblock {Formation of Dust Grains in Circumstellar Envelopes of Oxygen-Rich
  AGB Stars}.
\newblock \emph{Astrophysics and Space Science} {251}, 165--170 (1997).

\bibitem[{Verhoelst} et~al.(2006){Verhoelst}, {Decin}, {van Malderen}, {Hony},
  {Cami}, {Eriksson}, {Perrin}, {Deroo}, {Vandenbussche}, and
  {Waters}]{Verhoelst2006}
{Verhoelst}, T. \emph{et al.}
\newblock {Amorphous alumina in the extended atmosphere of {$\alpha$} Orionis}.
\newblock \emph{Astron. Astrophys.} {447}, 311--324 (2006).



\end{thebibliography}

\bigskip

\section*{Acknowledgements}
Based on observations collected with the NACO instrument at the European Southern Observatory, Chile.

\smallskip
\noindent
Correspondence and requests for materials should be addressed to bnorris@physics.usyd.edu.au

\end{document}